\documentclass{article}

\def\be{\begin{equation}}
\def\ee{\end{equation}}
\def\bea{\begin{eqnarray}}
\def\eea{\end{eqnarray}}
\def\({\left(}
\def\){\right)}

\begin{document}

\pagestyle{empty}
\vskip-10pt
\vskip-10pt
\hfill {\tt hep-th/0310037}
\begin{center}
\vskip 3truecm
{\Large\bf
On the Weyl anomaly of Wilson surfaces
}\\ 
\vskip 2truecm
{\large\bf
Andreas Gustavsson
}\\
\vskip 1truecm
{\it Institute of Theoretical  Physics,
Chalmers University of Technology, \\
S-412 96 G\"{o}teborg, Sweden}\\
\vskip 5truemm
{\tt f93angu@fy.chalmers.se}
\end{center}
\vskip 2truecm
\noindent{\bf Abstract:} 
We compute the Weyl anomaly for an abelian Wilson surface by using a regularization that respects the gauge invariance. We then study the loop space on which lives a one-form connection. We restrict ourselves to the subsector consisting of only concentrical circular loops, and derive a Maxwell type action on this restricted loop space.
\vfill \vskip4pt

\eject
\newpage
\pagestyle{plain}
 
\section{Introduction}
In this letter we compute the Weyl anomaly for surface observables in the quantum theory that arise upon quantizing an abelian two-form gauge field potential $B_{\mu\nu}$ in six euclidean dimensions. In the last section we also make an attempt to work in loop space in which the Wilson surface becomes a Wilson line (or Wilson loop if we had used loops with a common base point). The point is that Wilson lines have a non-abelian generalization. One problem here is of course that the loop space generically is infinite dimenional. We will in this letter only restrict ourselves to a finite dimensional subsector of the loop space.

The vacuum expectation value of an abelian Wilson surface is divergent. In \cite{HS} it was renormalized in a way that was not obviously gauge invariant. The regularization cut holes in the Wilson surface, and the abelian Wilson surface is gauge invariant only for closed surfaces. Furthermore that regularization can not be taken over to Minkowski signature because a small distance measured in Minkowski signare is not a compact set but is spread all around the vicinity of the light cone. We will here use a different regularization which respects the gauge invariance, and which can be taken over to Minkowski signature. The classical abelian action that we will start from reads
\be
S_0 = -\frac{1}{12}\int d^6 x \sqrt{G} H_{\mu\nu\rho}H^{\mu\nu\rho}\label{s0}
\ee
where $H_{\mu\nu\rho} = 3\partial_{[\mu}B_{\nu\rho]}$. From this action action we can not directly compute a propagator due to the gauge invariance. Imposing the covariant gauge condition $D_{\mu}B^{\mu\nu}-\omega^{\nu}(x)=0$ we can then use the 't Hooft trick and avarage over $\omega$ to get the gauge fixing term 
\be
S_{gf} = -\frac{\alpha}{2}\int d^6 x \sqrt{G}(D_{\mu}B^{\mu\nu})^2\label{sgf}
\ee
that we thus add to the classical action. Here $\alpha$ is an arbitrary non-zero parameter. We also should add a term for two anticommuting vector ghosts and terms for three commuting scalar ghosts for ghosts \cite{Siegel}. But these decouple in the abelian theory. If we add the gauge fixing term to the action we can compute the propagator, which then of course will depend on $\alpha$. When one computes a gauge invariant quantity, this quantity must not depend on the parameter $\alpha$. It does however not seem to be so easy to compute the Weyl anomaly for any other choice of $\alpha$ than for $\alpha=1$. So we can not keep $\alpha$ arbitrary throughout the computation to show that the end result really is independent of $\alpha$. 

The vacuum expectation value of an abelian Wilson surface $W(\Sigma) = \exp i\int_{\Sigma}B$ may be expressed as
\be
\left<W(\Sigma)\right> = e^{-I}
\ee
where
\be
I = \int_{(X,Y)\in \Sigma\times\Sigma} \Delta(X,Y).
\ee
Here $\Delta$ is the gauge field propagator. The integral is singular on the diagonal, that is when $X=Y$. We may regularize the divergence and then get
\be
I_{\epsilon} = \epsilon^{-2}I_2 + \log \epsilon I_0 + I_{fin} + Ordo(\epsilon).
\ee
The divergences here are local expressions integrated over $\Sigma$ and may therefore be removed by local counterterms in the action. The Weyl anomaly ${\cal{A}}$ is given by the conformal variation of the renormalized Wilson surface $W_{ren}(\Sigma) = e^{-I_{fin}}$ as,
\be
\delta W_{ren}(\Sigma) = {\cal{A}}(\Sigma) W_{ren}(\Sigma).
\ee
In order for $W_{ren}$ to have any physical significance it must of course be gauge invariant. But if one uses a regularization which is not manifestly gauge independent, this is not obviously the case.

Let us see how different choices of the gauge parameter $\alpha$ could change the result obtained in \cite{HS} at most. To keep things as simple as possible we make an infinitesmal conformal transformation around a flat background metric
\be
G_{\mu\nu} = \delta_{\mu\nu}.
\ee
In the covariant gauge above the gauge propagator is given by
\be
\overline{\triangle}_{\mu\nu}{}^{\rho\sigma}(x,y) = \frac{\delta_{\mu\nu}^{\rho\sigma}}{|x-y|^4} + \xi \delta_{[\mu}^{[\rho} \frac{(x-y)^{\sigma]}(x-y)_{\nu]}}{|x-y|^6}.\label{propagator}
\ee
where $\xi$ depends on $\alpha$ as $2\xi - 4\eta + \eta \xi = 0$ where $\eta = \alpha - 1$. If we make a conformal variation $G_{\mu\nu} \rightarrow e^{2\phi}\delta_{\mu\nu}$ we get to linear order in $\phi$ the short distance behaviour (we still rise indices by $\delta^{\mu\nu}$)
\be
\triangle_{\mu\nu}{}^{\rho\sigma}(x,y) = \overline{\triangle}_{\mu\nu}{}^{\rho\sigma}(x,y) 
+\omega \delta_{[\mu}^{[\rho}\partial_{\nu]}\phi(y)\frac{(x-y)^{\sigma]}}{|x-y|^4}
+\kappa \partial_{\lambda}\phi(y) \delta_{[\mu}^{[\rho}\frac{(x-y)^{\sigma]}(x-y)_{\nu]}(x-y)^{\lambda}}{|x-y|^6}+...
\ee
Here $\omega$ and $\kappa$ are coefficients that depend on $\alpha$, i.e. on the gauge choice. For the gauge choice $\alpha = 1$ made in \cite{HS} we get $\kappa = 0$ and $\xi=0$. For other gauge choices the $\kappa$-term will not vanish. Since $\partial_{\mu}\phi$ enters in all the gauge dependent terms (the $\omega$-term and the $\kappa$-term), we deduce that terms in the Weyl anomaly which do not involve derivatives of $\phi$ are always gauge independent. But if the regularization is not gauge invariant, then terms which involve derivatives of $\phi$ may be gauge dependent. 

We now show explicitly that the regularized quantity $I_{\epsilon}$ is not gauge invariant if one uses the regularization in \cite{HS}. If we assume that $D$ is a surface which has as its boundary $\Sigma$, then by using the Stokes theorem we get
\bea
I_{\epsilon} &=& \int_{\Sigma\times\Sigma} \Delta(X,Y)\theta(s^2(X,y)-\epsilon^2)\cr
&=& \int_{X\in\Sigma}\int_{y\in D} d_y \(\Delta(X,y)\theta(s^2(X,y)-\epsilon^2)\).\label{reg}
\eea
Acting with the exterior derivative on the propagator yields a piece that is gauge invariant. But when we hit the step function we get a term which apparently is not gauge invariant. It is given by
\be
\int_{\Sigma} dX^{\mu}\wedge dX^{\nu} \int_{D}
dy^{\sigma}\wedge dy^{\rho}\wedge dy^{\lambda} \Delta_{\mu\nu,\sigma\rho}(X,y)\partial_{\lambda}\theta(s^2(X,y)-\epsilon^2).\label{gauge}
\ee
We may compute the integral over $D$ by choosing Riemann normal coordinates around $X$. We denote them $y^{\mu}$, though in these coordinates we let the point $X$ lie at the origin $y=0$. According to \cite{HS} we then have $s^2(X=0,y) = |y|^2  + O((y)^5)$. No restriction is made by assuming that the tangent plane of $\Sigma$ in $X$ is the $1,2$-plane, and we may choose $D$ to be the $1,2,3$-semispace outside (or above) $\Sigma$, at least in the vicinity of $X$. Introducing spherical coordinates, ($d\Omega$ denotes the surface element on the unit two-sphere and $y$ is a radius) we get the integral over $D$ as
\be
\int_D dy d\Omega y^2 \Delta_{12,[12}(y)2y_{3]}\frac{\delta(y-\epsilon)}{2y} = \epsilon \int_{S^2_{\epsilon}(X) \cap D}d\Omega \Delta_{12,[12}(y)y_{3]}.
\ee
In flat space this yields something that goes like $\epsilon^{-2}$. Noting (\ref{propagator}) it also very easy to realize that it also depends on the gauge parameter $\alpha$. In a curved background we also get $\epsilon^{-1}$ divergences as well as finite terms (plus terms which tend to zero as $\epsilon$ tends to zero), which all involve derivatives of the metric tensor. The $\epsilon^{-1}$ divergence depends on the choice of $D$ and must be cancelled by an opposite such divergence in the gauge invariant piece of $I_{\epsilon}$ that also depends on $D$ since the sum of these two terms is independent of the choice of $D$ (as evident from (\ref{reg})). 

Under a conformal variation of $I_{\epsilon}$ the finite contributions involve derivatives of $\phi$. Dimensional analysis and covariance (independence of the choice of $D$) alone implies that the term that involves one derivative of $\phi$ must be proportional to the last term in equation (\ref{anomaly}). We may notice that it is impossible for the gauge dependent terms to depend on the choice of $D$. (In particular we deduce that the $\epsilon^{-1}$ divergences must actuallt be gauge independent. Of course some, but not all, individual terms in (\ref{gauge}) may be independent of the gauge parameter $\alpha$). If one adds the gauge invariant piece of $\delta I_{\epsilon}$ (here $\delta$ is a conformal variation) we get something that does not depend on $D$. Now if we change the gauge-parameter $\alpha$, we expect only the {\sl{coefficients}} of the terms in the gauge dependent piece of $\delta I_{\epsilon}$ to change. Taking the difference of these two $D$-independent expressions (the one before and the one after the gauge variation), we get an expression that only involves the gauge dependent terms (though with different coefficients), and the difference of two $D$-independent expressions is of course still $D$-independent. We have not proved that finite gauge dependent terms really are there, but we may give one convincing argument for their existence. If one considers the curvature (or the second fundamental form) of the embedded surface $\Sigma$ at $X$ and compare this to a (locally) flat surface $\Sigma'$ at $X$ we get different domains of integration $S^2_{\epsilon}(X) \cap D$. Therefore the difference between $\delta I_{\epsilon}$ before and after such a deformation of $\Sigma$ should involve terms that depend on the second fundamental form. On dimensional grounds such terms must be finite or tend to zero as $\epsilon$ tends to zero. They should also be independent of $D$ since the difference only touch the area near the boundary of $S^2_{\epsilon}(X) \cap D$, and so should only depend on (local) properties of $\Sigma$.

The above mentioned finite gauge dependent term is a Type $D$ anomaly, using the terminology in \cite{HS}. This is a Type $C$ anomaly which is integrated only over a submanifold, which implies that there is no counterterm that can remove the type $D$ anomaly, as argued in \cite{HS}. It is thus part of the Weyl anomaly and so it is essential to determine its coefficient. In principle this coefficient should be possible to determine by first computing the conformal variation of the gauge dependent counterterms that has to be added to $I_{\epsilon}$ in order to make this quantity gauge invariant (thereby removing any artificial gauge anomaly), and then add this contribution to the result obtained in \cite{HS}. However this computation does not seem to be so easy to carry out in practice. The difficulty lies in rewriting the gauge-dependent $D$-independent terms in $I_{\epsilon}$ as local integrals over $\Sigma$, and then to compute their conformal variations. 

In this letter we will instead use a regularization which is manifestly gauge invariant from start and redo the computation of the Weyl anomaly in order to determine the numerical value of this coefficient. We will not get the same numerical value as in \cite{HS}, which further indicates that their renormalization perhaps was not really gauge invariant.

\section{Regularization of the Wilson surface observable}

We will choose Riemann normal coordinates around a point $X$ in $\Sigma$ and let
\be
G_{\mu\nu}(X)=\delta_{\mu\nu}
\ee
in that point. The curvature is not necessarily zero at $X$. The second derivatives of the metric (and hence the curvature tensor) may be non-zero at $X$, but all first order derivatives of the metric tensor vanishes at $X$ in Riemann normal coordinates.

We can always find a local Riemann normal coordinate system in which the Wilson surface is located at a point in say the $x^0$ direction. We then regularize the Wilson surface by translating one of the surfaces $\Sigma$ to $\Sigma'$ a distance $\epsilon$ in the $x^0$ direction. Particular care then has to be taken when one goes from one coordinate patch to another, but such global issues will not concern us here since the Weyl anomaly is a local quantity. We thus have that
\be
X^0(Y^i) = X^0(X^i)
\ee
for any two points $X^{i}$ and $Y^{i}$ on $\Sigma$ ($i=1,...,5$), whereas
\be
Y^0(Y^i) = X^0(Y^i) + \frac{\epsilon}{\sqrt{G_{00}(Y^i)}} = X^0(X^i) + \frac{\epsilon}{\sqrt{G_{00}(Y^i)}}
\ee
generically will depend on $Y^i$ through the zero-zero component of the metric tensor. Assuming Riemann normal coordinates around $X$ we do not get any first order correction when Taylor expanding the metric,
\be
G_{00}(Y) = G_{00}(X) + Ordo(|Y^i-X^i|^2) = 1 + Ordo(|Y^i-X^i|^2).
\ee
Now we will also be interested in the conformal variation of this. After a conformal variation the metric becomes
\be
e^{2\phi(Y)}G_{00}(Y) = e^{2\phi(X) + (Y-X)^i \partial_i \phi(X)} + Ordo(|Y-X|^2).
\ee

In principle we have to compute the conformal variation of $I_{fin}$. One might then think that one must first compute $I_{fin}$. But that is in general difficult (if not impossible) to do for a generic surface $\Sigma$. Fortunately one does not have to compute $I_{fin}$ first. We first notice that the variations of $I_2$, $I_0$ and $I_{fin}$ all can be extracted by making a variation of $I_{\epsilon}$ while keeping $\epsilon$ fixed.\footnote{This is merely a mathematical trick in order to extract $\delta I_{\epsilon}$. Physically $\epsilon$ has the dimension of a length and should also vary under a conformal variation.} Choosing Feynman gauge ($\alpha=1$) we will find that one does only have to compute $I_0$ (the coeffiecient of the logarithmic divergence) in order to compute $\delta I_{fin}$, and the divergent terms are much easier to compute as they depend only on the local properties of $\Sigma$ in contradistinction to $I_{fin}$ which depends on properties of the whole surface $\Sigma$.

In Feynman gauge and in Riemann normal coordinates, the gauge field propagator and its variation was computed in \cite{HS} to
\be
\Delta[e^{2\phi}G_{\mu\nu}](X,Y)_{ij}^{kl} - \Delta[G_{\mu\nu}](X,Y)_{ij}^{kl} = {\mbox{exact form}}
\ee
where 
\bea
\Delta[G_{\mu\nu}](X,Y)_{ij}^{kl} &=& \frac{\delta_{ij}^{kl}}{|X-Y|^4} \cr
&+& \(\frac{4}{3}P_{[i}{}^{[k}\delta_{j]}^{l]}-\frac{1}{2}W_{[i}{}^{[k}{}_{j]}{}^{l]}\)\frac{1}{|X-Y|^2}\cr
&-&
\frac{1}{3}\(P^{[k}{}_{m}\delta_{n[i}\delta_{j]}^{l]} + \delta^{[k}{}_{m}P_{n[i}\delta_{j]}^{l]} + W^{[k}{}_{mn[i}\delta_{j]}^{l]}\)\frac{(X-Y)^m(X-Y)^n}{|X-Y|^4}
\eea
up to a normalization constant $-1/(4V_5)$ where $V_5$ is the volume of a unit five-sphere, which we will leave out. Here $P_{\mu\nu}=\frac{1}{4}\(R_{\mu\nu}-\frac{1}{10}RG_{\mu\nu}\)$.

Then the conformal variation of $I_{\epsilon}$ with $\epsilon$ kept {\sl{fixed}} is given by
\bea
\delta I_{\epsilon} & = & 
\int_{X^0} \int_{Y^0(Y) = X^0(X) + \frac{\epsilon}{e^{\phi(Y)}}} 
\Delta[e^{2\phi}G_{\mu\nu}](X,Y) - \int_{X^0} \int_{Y^0(Y) = X^0(X) + \epsilon}\Delta[G_{\mu\nu}](X,Y)\cr
& = & \int_{X^0} \int_{Y^0(Y) = X^0(X) + \frac{\epsilon}{e^{\phi(Y)}}} 
\(\Delta[e^{2\phi}G_{\mu\nu}](X,Y) - \Delta[G_{\mu\nu}](X,Y)\)\cr
&& + \(\int_{X^0} \int_{Y^0(Y) = X^0(X) + \frac{\epsilon}{e^{\phi(Y)}}}\Delta[G_{\mu\nu}](X,Y)
 - \int_{X^0} \int_{Y^0(Y) = X^0(X) + \epsilon}\Delta[G_{\mu\nu}](X,Y)\)
\eea
The first double integral is over an exact form and hence that whole integral vanishes and we are left with
\bea
\delta I_{\epsilon} & = & \int\int \left\{\Delta[G_{\mu\nu}]\(\frac{\epsilon}{e^{\phi(Y)}};X^i,Y^j\) - \Delta[G_{\mu\nu}](\epsilon;X^i,Y^j)\right\}\cr
& = & -\epsilon \frac{\partial}{\partial \epsilon}\int\int \(\phi(X) + (Y-X)^i \partial_i \phi(X) + ...\)\Delta[G_{\mu\nu}](\epsilon;X^i,Y^j).
\eea
Noticing that $\epsilon \frac{\partial}{\partial \epsilon} = \frac{\partial}{\partial \log \epsilon}$ we thus see that the conformal variation does not have any logarithmic divergence, only a quadratic divergence, and that the anomaly can be read off as (minus) the coefficent of the logarithmic divergence (as the quadratic divergence can be absorbed by a covariant counterterm \cite{HS}).

\section{The Weyl anomaly}
We thus want to compute the logarithmic divergence of the integral
\be
\int_{\Sigma} dX^i\wedge dX^j \int_{\Sigma} dY^k \wedge dY^l \(\phi(X) + (Y-X)^i \partial_i \phi(X) + ...\)\Delta_{ij,kl}[G_{\mu\nu}](\epsilon;X^i,Y^j).
\ee
This we do by first fixing a point on $\Sigma$ and choosing Riemann normal coordinates around this point such that the tangent plane to this point is the $X^1,X^2$-plane and such that the point itself is at $X^1=X^2=0$ in this tangent plane. The second integral will receive its divergent part only from a small vicinity around this point. The behaviour of $\Sigma$ far away from this point does not affect the coefficient of the logarithmic divergence $\log \epsilon$. 

In a flat background (we will consider the contributions that arise when the background is curved in a moment) we only have to consider the following two integrals,
\be
A = \int_{\Sigma} dY^1\wedge dY^2 \frac{1}{(\epsilon^2 + |X^i-Y^i|^2)^2}
\ee
and
\be
D = \int_{\Sigma} dY^1\wedge dY^2 \frac{(Y-X)^3}{(\epsilon^2 + |X - Y|^2)^2}.
\ee
The idea now is to use the cartesian coordinates $Y^1,Y^2$ in the tangent plane $T_X \Sigma$. To compute the logarithmic divergence it will turn out to be enough just to keep the first few terms in an expansion like
\be
X^{\mu}(Y) = X^{\mu}(0) + Y^i \partial_i X^{\mu}(0) + \frac{1}{2}Y^i Y^j \partial_i\partial_j X^{\mu}(0)+...
\ee
Up to linear part we are still in $T_X \Sigma$, so we deduce that
\be
(Y - X)^{\mu} = \frac{1}{2}\Omega^{\mu}_{ij}Y^i Y^j + ...
\ee
where $\Omega^{\mu}_{ij} \equiv \partial_i\partial_j X^{\mu}(0)$. This is a symmetric matrix and hence we can rotate the tangent space coordinates $Y^i$ so that it becomes diagonal. In the sequel we will assume this has been done. A coordinate in the tangent plane is denoted by $Y^i$ (i=1,2) whereas a coordinate on $\Sigma$ by $Y^{\mu}$. Using the Pythagoras theorem, $|Y^{\mu}(Y^i)-X^{\mu}|^2 = |Y^i|^2 + \frac{1}{4}|\Omega^{\mu}_{ij} Y^i Y^j|^2 + ...$ we get
\bea
A & = & \int dY^1\wedge dY^2 \frac{1}{(\epsilon^2 + |Y^i|^2 + \frac{1}{4}|\Omega^{\mu}_{ij} Y^i Y^j|^2 + ...)^2}\cr
& = & \int_0^{2\pi} d\varphi \int_0^{?} dY Y \frac{1}{\(\epsilon^2 + Y^2 + \frac{1}{4}\(\Omega^{\mu}(\varphi)\)^2 Y^4+...\)^2}
\eea
where $\Omega^{\mu}(\varphi) \equiv \Omega^{\mu}_{11} \cos^2 \varphi + \Omega^{\mu}_{22} \sin^2 \varphi$. The divergent piece of an integral like $\int_0^{?} \frac{du}{(\epsilon^2 + u + \alpha u^2 + \beta u^3 + ...)^2}\equiv f(\alpha,\beta,...)$ can be extracted by Taylor expanding $f$ around $\alpha=\beta=...=0$ as $f(\alpha,...) = f(0) + \alpha f'_{\alpha}(0) + ...$. Here $f(0)$ goes like $\epsilon^{-2}$ and $f'_{\alpha}(0)$ like $4\log \epsilon$ and all higher order terms in the Taylor expansion are finite or goes to zero as $\epsilon$ goes to zero. Applying this to the integral above we get the  divergent piece as
\bea
&&\int_0^{2\pi}d\varphi \frac{1}{2}\(\epsilon^{-2} + (\Omega(\varphi))^2 \log \epsilon\)\cr
&&=\pi\epsilon^{-2} + \frac{1}{2} \(\frac{3\pi}{4}(\Omega_{11})^2+\frac{\pi}{2}\Omega_{11}\cdot \Omega_{22}+\frac{3\pi}{4}(\Omega_{22})^2\)\log \epsilon\cr
&&=\pi\epsilon^{-2} + \frac{1}{8}\({\pi}\(\Omega_{ij}\delta^{ij}\)^2 + {2\pi}\Omega_{ij}\Omega_{kl}\delta^{ik}\delta^{jl}\)\log \epsilon
\eea
Similarly we compute
\bea
D & = & \int dY^1\wedge dY^2 \frac{(Y-X)^3}{(\epsilon^2 + |X - Y|^2)^2}\cr
& = & \int d\varphi \int dY Y \frac{\frac{1}{2}\Omega^3(\varphi)Y^2}{(\epsilon^2 + Y^2 +...)^2}\cr
& = & -\frac{1}{2}\pi \Omega^3_{ij}\delta^{ij} \log \epsilon + {\mbox{finite terms}}.
\eea

To compute the anomaly in a curved background we need also the following two integrals,
\bea
A' & = & \int dY^1\wedge dY^2 \frac{1}{\epsilon^2 + |X-Y|^2} = -2\pi\log \epsilon + {\mbox{finite terms}}\cr
A'' & = & \int dY^1\wedge dY^2 \frac{\beta_{ij}Y^i Y^j}{(\epsilon^2 + |X-Y|^2)^2} = -\pi \beta_{ij}\delta^{ij} \log \epsilon + {\mbox{finite terms}}.
\eea
Then we get the contribution
\bea
&&\int dY^1\wedge dY^2 \left\{ \(\frac{4}{3}P_{[1}{}^{[1}\delta_{2]}^{2]}-\frac{1}{2}W_{[1}{}^{[1}{}_{2]}{}^{2]}\)\frac{1}{|X-Y|^2}\right.\cr
&&\left.+\frac{1}{3}\(P^{[1}{}_{m}\delta_{n[1}\delta_{2]}^{2]} + \delta^{[1}{}_{m}P_{n[1}\delta_{2]}^{2]} + W^{[1}{}_{mn[1}\delta_{2]}^{2]}\)\frac{(X-Y)^m(X-Y)^n}{|X-Y|^4}\right\}\cr
&&=-\frac{1}{2}\(\pi P_{ij}\delta^{ij} - \frac{\pi}{3}W_{ikjl}\delta^{ij}\delta^{kl}\)\log \epsilon + {\mbox{finite terms}}.
\eea

We want to write the Weyl anomaly in an arbitrary coordinate system, and so we need to find expressions that are covariant under diffeomorphisms of the six dimensional target space as well as under reparametrizations $X^i \mapsto \sigma^{\alpha} = \sigma^{\alpha}(X^j)$ of the Wilson surface, and which reduce to the terms found above in the particular coordinate system that we used there. The matrices $\Omega^{\mu}_{ij}$ uniquely covariantizes to the second fundamental form $\Omega^{\mu}_{\alpha\beta}$. The Kronecker delta $\delta_{ij}$ covariantizes to the induced metric $g_{\alpha\beta}$ on $\Sigma$. 

Summing up, the Weyl anomaly is given by the covariantized expression
\bea
 \left[\frac{1}{2}\(-\frac{\pi}{8} (g^{\alpha\beta}\Omega_{\alpha\beta})^2 
-\frac{\pi}{4}\Omega_{\alpha\beta}\Omega_{\gamma\delta}g^{\alpha\gamma}g^{\beta\delta}
+ \frac{1}{2}\pi g^{\alpha\beta}\Omega^{\mu}_{\alpha\beta}D_{\mu}\) 
+\frac{1}{2}\(\pi P_{\alpha\beta}g^{\alpha\beta} 
-\frac{\pi}{3}W_{\alpha\beta\gamma\delta}g^{\alpha\gamma}g^{\beta\delta}\)\right]\phi.
\eea
The additional factor of $\frac{1}{2}$ came from $\delta_{12}^{12} = \frac{1}{2}$ in the 'flat space part' of the propagator. Using the identity,
\be
\Omega_{\alpha\beta}\Omega_{\gamma\delta}g^{\alpha\gamma}g^{\beta\delta} 
= (g^{\alpha\beta}\Omega_{\alpha\beta})^2 + R_{(2)} - g^{\alpha\gamma}g^{\beta\gamma}W_{\alpha\beta\gamma\delta} 
- 2g^{\alpha\beta} P_{\alpha\beta}
\ee
which is a generalization of the Gauss-Bonnet theorem to a two-dimensional surface embedded in a curved background, we find that the Weyl anomaly may be written as
\be
\frac{\pi}{4}\left\{-\(\frac{3}{4} \((g^{\alpha\beta}\Omega_{\alpha\beta})^2 - 4 g^{\alpha\beta}P_{\alpha\beta}\) + \frac{1}{2} R_{(2)} + \frac{1}{6}g^{\alpha\gamma}g^{\beta\delta} W_{\alpha\beta\gamma\delta}\)\phi + g^{\alpha\beta}\Omega^{\mu}_{\alpha\beta} D_{\mu}\phi\right\}.\label{anomaly}
\ee
The similarity of this result with that of \cite{HS} was explained already in the introduction.

In a flat background this Weyl anomaly reduces to
\be
\frac{\pi}{4}\left\{-\(\frac{3}{4} (g^{\alpha\beta}\Omega_{\alpha\beta})^2 + \frac{1}{2} R_{(2)}\)\phi +g^{\alpha\beta}\Omega^{\mu}_{\alpha\beta}\partial_{\mu}\phi\right\}
\ee
This is in general not proportional to the Weyl anomaly in the large $N$ limit which in a flat spacetime reduces to something that is proportional to \cite{GW}
\be
\frac{1}{2}(g^{\alpha\beta}\Omega_{\alpha\beta})^2\phi - g^{\alpha\beta}\Omega^{\mu}_{\alpha\beta}\partial_{\mu}\phi.
\ee
An exception is if the Wilson surface is taken to be a round sphere in flat ${\bf{R}}^6$ with metric $G_{\mu\nu}=\delta_{\mu\nu}$. For a unit sphere, $R_{(2)}=-2$ and $\Omega^3_{ij} = - \delta_{ij}$, and in this case we indeed find agreement of the Weyl anomalies.

In this case we can actually compute the abelian Wilsons surface exactly. Parametrizing the spherical Wilson surface as
\bea
x^1 & = & \sqrt{1-Y^2}\cos{\rho}\cr
x^2 & = & \sqrt{1-Y^2}\sin{\rho}\cr
x^3 & = & Y,
\eea
we can compute $I_{\epsilon}$ by computing the integral over the second surface $\Sigma=S^2$ while keeping any point $X$ on the first surface fixed, and then multiply by the volume $V_2$ of the two-sphere. Due to rotational invariance of the sphere we will get the same result irrespectively of which point we choose on the first surface. But the second integral simplifies mostly if we choose one of the poles, say the north pole. We then get\footnote{This result was first obtained by E. Flink in his masters thesis.}
\bea
I_{\epsilon} &=& V_2\frac{1}{2}\int dY\wedge d\rho
\frac{Y}{\(\epsilon^2 + 2 - 2 Y \)^2}\cr
&=& V_2\frac{\pi}{2} \(\frac{1}{\epsilon^2} + \log \epsilon - \log 2 + \frac{1}{4} + O(\epsilon)\).
\eea
The divergent piece is what the general formalism above also would give. The scale anomaly is the coefficient of $-\log{\epsilon}$.

It would of course be very interesting to understand why the Weyl anomaly coincides with the result for large $N$ for a sphere. We hope to be able to explain this in a future work by making a $1/N$ expansion on the gauge theory side. One way to compute the non-abelian Wilson surface might be to work in loop space which we will discuss briefly now.

\section{Loop space}
An abelian two-form connection on a six-dimensional base manifold X may alternatively be formulated as a one-form connection on loop space ${\cal{L}}X$ as \cite{Hofman}
\be
A(\gamma) = \int_0^{2\pi} d\sigma B_{\mu\nu}(X(\sigma))dX^{\mu}(\sigma)\frac{\partial X^{\nu}}{\partial \sigma},
\ee
where the loop $\gamma$ is parametrized by $\sigma\in [0,2\pi]$. This definition does not quite work in the non-abelian case however because we are adding tensors at different points in $X$. A proper definition would require a one-form connection on $X$ in order to be able to parallel transport all the tensors to one and the same point where we then can add them together. In reference \cite{Hofman} such a non-abelian generalization has been done. 

Here we will instead do explicit computations in a however very restricted loop space which we denote ${\cal{L}}_C X$ which consists of only circular loops in the $x^{1,2}$ plane centered at $x^1=x^2=0$ in a flat base manifold $X={\bf{R}}^6$.

The propagator for the one-form connection in ${\cal{L}}_C X$ can be obtained directly from the propagator in $X$ as
\be
\left<A_{m}(\gamma)A_{n}(\delta)\right> = \int_{\gamma} d\sigma \int_{\delta} d\rho 
\frac{\partial X^{\mu}}{\partial \sigma}
\frac{\partial X^{\nu}}{\partial X^m}\frac{\partial Y^{\kappa}}{\partial \rho}
\frac{\partial Y^{\tau}}{\partial Y^n} \left<B_{\mu\nu}(X)B_{\kappa\tau}(Y)\right>.
\ee
Here we used $x^m=(x^0,r,x^3,x^4,x^5)$ as coordinates on ${\cal{L}}_C X$, specifying the 'center of mass' and radius of the loop (i.e. of the circle in the $x^{1,2}$-plane) in $X$. If we as our loops take
\bea
\gamma: \sigma & \mapsto & (X^0,r\cos\sigma,r\sin\sigma,X^3,X^4,X^5)\cr
\delta: \rho & \mapsto & (Y^0,s\cos\rho,s\sin\rho,Y^3,Y^4,Y^5),
\eea
we get (we will let $M,N = 0,3,4,5$)
\bea
\left<A_{M}(\gamma)A^{N}(\delta)\right>
&=& \delta_M^N rs\int_{\gamma} d\sigma \int_{\delta} d\hat\rho
\cos\hat\rho \frac{1}{(|X^M-Y^M|^2+r^2+s^2-2rs\cos\hat\rho)^2}\cr
\left<A_{r}(\gamma)A^{r}(\delta)\right>
&=& rs\int_{\gamma} d\sigma \int_{\delta} d\hat\rho\frac{1}{(|X^M-Y^M|^2+r^2+s^2-2rs\cos\hat\rho)^2}.
\eea
all other components vanishing. These definite integrals can be evaluated explicitly using that
\be
\int_0^{2\pi} d\rho \frac{1}{1+a\cos \rho} = \frac{2\pi}{\sqrt{1-a^2}}.
\ee
We then get the propagator on ${\cal{L}}_C X$ as (up to factors of $2\pi$)
\bea
\left<A_M(\gamma)A^N(\delta)\right> &=& \delta_M^N\frac{(rs)^2}{\(\(|x^M-y^M|^2+r^2+s^2\)^2-4r^2s^2\)^{3/2}}\cr
\left<A_r(\gamma)A^r(\delta)\right> &=& \frac{rs\(|x^M-y^M|^2+r^2+s^2\)}{\(\(|x^M-y^M|^2+r^2+s^2\)^2-4r^2s^2\)^{3/2}}.\label{prop}
\eea
 
We can also derive this propagor from an action on ${\cal{L}}_C X$. This action can be derived from (\ref{s0}) and (\ref{sgf}) by introducing planar polar coordinates $x^1 = r \cos \sigma$ and $x^2 = r \sin \sigma$ on $X$ and imposing rotational symmetry $\partial_{\sigma}=0$ as a constraint already at the level of that action. Then the loop space connection $A_{m}(\gamma)$ will be proportional to $B_{m\sigma}$ evaluated at any point $x$ on the loop (cirle) $\gamma$. If we decompose $\mu = (m,\sigma)$ where in turn $m = (0,r,3,4,5)$ we find that the components $B_{mn}$ decouple from $B_{m\sigma}$. We can thus study the theory for $A_{m}$ separately. Its action becomes\footnote{We impose the boundary conditions $\partial_r|_{r=0} = 0$ since we do not expect any cusp singularity at the origin and assume that the fields go to zero sufficiently fast at infinity. Then we can make the integration by parts without getting any boundary terms.}
\be
S_0 = -\int d^4x dr \frac{1}{4r}F_{mn}F^{mn}
\ee
where $F_{mn} = \partial_m A_n - \partial_n A_m$, and the gauge fixing part is given by
\be
S_{gf} = -\int d^4x dr \frac{r}{2}\(\partial_m(r^{-1}A^m)\)^2.
\ee
from which we derive the Schwinger-Dyson equations
\bea
\(r^{-1}(\partial_{M})^2+r^{-1}(\partial_{r})^2-r^{-2}\partial_r\)\left<A_M(x,r)A^N(y,s)\right>&=&\delta_M^N\delta(x-y,r-s)\cr
\(r^{-1}(\partial_{M})^2+r^{-1}(\partial_{r})^2-r^{-2}\partial_r+r^{-3}\)\left<A_r(x,r)A^r(y,s)\right>&=&\delta(x-y,r-s)
\eea
Using {\sl{Mathematica}} we found that the propagotor (\ref{prop}) indeed does satisfy these equations.

The natural non-abelian generalization of the classical action is
\be
S_0 = -\int d^4x dr \frac{1}{4r} Tr(F_{mn}F^{mn})
\ee
where $F_{mn} = \partial_m A_n - \partial_n A_m + i[A_m,A_n]$
and $A_m = A^a_m T^a$ where $T^a$ are generators of a Lie group. This could be a starting point for computing the non-abelian Wilson surface for the class of surfaces that can be built up of concentrical circles, such as a two-sphere. We have not proved that if we span the surface with different loops we get the same result. A complication is that if we span the surface with different loops we get different actions on the loop space.

\vskip 0.5truecm
{\sl{Acknowledgments:}}

I would like to thank M Henningson for dicussions concerning the regularization and the second fundamental form and J Sikkema for making me aware of reference \cite{Hofman}.

\end{document}